\documentclass[preprint,letterpaper,showpacs,preprintnumbers,amsmath,amssymb,superscriptaddress,nofootinbib]{revtex4}
\def\figsize{10cm} 
\def\parasec#1{\section{#1}}
\def\introsec#1{\section{#1}}

\usepackage{graphicx}
\usepackage{dcolumn}
\usepackage{bm}

\usepackage{color}

\def\epspdffile#1{
  \includegraphics[height=\figsize]{#1.eps}
}

\begin{document}

\title{Adaptive multigrid algorithm for the lattice Wilson-Dirac operator}

\author{R.~Babich}
\affiliation{Center for Computational Science, Boston University,
3 Cummington Street, Boston, MA 02215, USA}
\affiliation{Department of Physics, Boston University,
590 Commonwealth Avenue, Boston, MA 02215, USA}
\author{J.~Brannick}
\affiliation{Department of Mathematics, The Pennsylvania State University,
 230 McAllister Building, University Park, PA 16802, USA}
\author{R.~C.~Brower}
\affiliation{Center for Computational Science, Boston University,
3 Cummington Street, Boston, MA 02215, USA}
\affiliation{Department of Physics, Boston University,
590 Commonwealth Avenue, Boston, MA 02215, USA}
\author{M.~A.~Clark}
\affiliation{Harvard-Smithsonian Center for Astrophysics, 60 Garden Street, Cambridge, MA 02138, USA}
\author{T.~A.~Manteuffel}
\affiliation{Department of Applied Mathematics, Campus Box 526, University of
Colorado at Boulder, Boulder, CO 80309, USA}
\author{S.~F.~McCormick}
\affiliation{Department of Applied Mathematics, Campus Box 526, University of
Colorado at Boulder, Boulder, CO 80309, USA}
\author{J.~C.~Osborn}
\affiliation{Argonne Leadership Computing Facility, Argonne National Laboratory, Argonne, IL 60439, USA}
\author{C.~Rebbi}
\affiliation{Center for Computational Science,  Boston University,
3 Cummington Street, Boston, MA 02215, USA}
\affiliation{Department of Physics, Boston University,
590 Commonwealth Avenue, Boston, MA 02215, USA}

\date{May 14, 2010}

\begin{abstract}
  We present an adaptive multigrid solver for application to the
  non-Hermitian Wilson-Dirac system of QCD. The key components leading
  to the success of our proposed algorithm are the use of an adaptive
  projection onto coarse grids that preserves the near null space of
  the system matrix together with a simplified form of the correction
  based on the so-called \(\gamma_5\)-Hermitian symmetry of the Dirac
  operator. We demonstrate that the algorithm nearly eliminates critical
  slowing down in the chiral limit and that it has weak dependence on
  the lattice volume.
\end{abstract}

\maketitle

\introsec{Introduction}


Perhaps the most severe computational challenge facing the lattice
approach to quantum chromodynamics is the divergent increase in cost
as one approaches the chiral limit required for the
experimental values of the up and down quark masses. (Similar difficulties
confront field theories conjectured for physics beyond the
standard model as well.)  The cause is well known: as the fermion mass
approaches zero, the Dirac operator becomes
singular (Re(\(\lambda_{\tiny{\mbox{min}}}\))\(\rightarrow0\)),
causing ``critical slowing down'' of the standard Krylov solvers
typically used to find the propagators. This is unavoidable for all
single-grid solvers.  Improving convergence with a suitable
preconditioning has been a main topic of research in lattice QCD for
many years but has, until recently, met very limited success in practice.

Eigenvector deflation~\cite{Stathopoulos:2007zi,Morgan:2007dq} is a
popular technique for accelerating solver convergence and is generally
successful provided sufficiently many eigenvectors are used in the
deflation process; exact deflation approaches are, however, expected
to scale as the square of the lattice volume \(O(V^2)\) and, thus,
become ineffective for large volumes.  An alternative is the local
deflation approach of \cite{Luscher:2007se}. 

Here approximate eigenvectors are used in the deflation process, and 
due to the local coherence (see below) of the low modes of the Dirac 
operator, only a volume-independent number of low-mode prototypes are 
required. As a result, an effective deflation of the operator is 
achieved with a computational effort growing approximately like $V$ 
rather than $V^2$.



Here we present an adaptive multigrid (MG) solver for the Dirac equation
\begin{equation} D(U)\psi = b \; , \end{equation}
where 
%

\begin{equation}
D_{x,y}(U)  =  (4 + m)\delta_{x,y}  -\sum^4_{\mu=1}[\frac{1-\gamma_\mu}{2} U^\mu_x\delta_{x+\hat\mu,y} + \frac{1+\gamma_\mu}{2}  U^{\mu\dagger}_{x-\hat\mu} \delta_{x-\hat\mu,y}]
\end{equation}
 is the Wilson lattice discretization of the Dirac operator.  This is expressed
(implicitly) as the tensor product of \(4\times4\) Dirac gamma
matrices $\gamma_\mu$ and \(3\times3\) $SU(3)$ gauge matrices
$U_\mu(x)$ on the nearest neighbor links $(x,y)$ of a hypercubic
spacetime lattice. While this matrix is not Hermitian, it satisfies
\(\gamma_5\)-Hermiticity (\(D^\dagger = \gamma_5 D \gamma_5\)); the
corresponding
Hermitian matrix, \(H = \gamma_5 D\), is maximally indefinite.  The
eigenvalues of \(D\) are complex and satisfy \(Re(\lambda_{\tiny{\mbox{min}}}) > 0\) for
physical values of the simulation parameters.

In a previous work~\cite{Brannick:2007ue}, we presented an algorithm
for solving the normal equations obtained from the Wilson-Dirac system in the
context of 2 dimensions, with a \(U(1)\) gauge field.  Here, we extend
this approach to directly solve the Wilson-Dirac system and apply the
resulting algorithm to the full 4-dimensional \(SU(3)\) problem.

\parasec{\bf Adaptive Multigrid}

The ``low'' modes, eigenmodes with small-in-magnitude eigenvalues of
the system matrix, are typically those responsible for the poor convergence
suffered by
standard iterative solvers (relaxation or Krylov methods).  As the
operator becomes singular, the error in the iteratively computed
solution quickly becomes dominated by these modes.  In the free field
theory, these slow-to-converge modes are geometrically smooth and,
hence, can be well represented on a coarse grid using fewer degrees of
freedom.  Moreover, these smooth modes on the fine grid now again
become rough (high frequency) modes on the coarse grid.  This
observation motivated the classical geometric MG approach, in which
simple local averaging is used to restrict residuals to the coarse
grid and linear interpolation is used to transfer corrections
(obtained from solving the coarse-grid error equation) to the fine grid.
We hereafter denote the interpolation operator by \(P\) and
restriction operator by \(R\).

Given a Hermitian positive definite (HPD) operator \(A\), taking the
restriction operator as \(R = P^\dagger\) and the coarse-grid operator
as \(A_c = P^\dagger A P\) gives the optimal (in an energy-norm sense)
two-grid correction.  It is natural to extend this recursively by
defining the problem on coarser and coarser grids until the degrees of
freedom have been reduced enough to permit an exact solve.  When
combined with \(m\) pre-relaxations (before restriction) and \(n\)
post-relaxations (after prolongation) on each level, we arrive at the
usual \(V(m,n)\)-cycle.  Such an MG process is known to eliminate
critical slowing down for discretized elliptic PDE problems, scaling
as \(O(V)\)~\cite{Brandt:1977}.

Explicitly, the error propagation operator for the two-grid solver with a single post-relaxation smoother \(S\) is given by
\begin{eqnarray}
E_{TG}&=&S(I - P(P^\dagger  A P)^{-1} P^{\dagger} A).
\end{eqnarray}
The performance of the MG algorithm is related to
\(\operatorname{range}(P)\) and how well this approximates the
slow-to-converge modes of the chosen relaxation procedure. Given a
convergent smoother, the two-grid algorithm can be shown to converge
(i.e., $\|E_{TG}\|_A<1$) provided that range(\(P\)) approximates eigenvectors
with error proportional to the size of their corresponding
eigenvalues.


For the Wilson-Dirac system in the interacting theory, the low modes
are not geometrically smooth, and so classical MG approaches, which
assume the slow-to-converge error is locally constant, fail
completely.  In such settings, the gauge field is essentially random
and causes local oscillations in the low modes.  Moreover, the
proceedure is not inherently gauge invariant and would require finding
a suitably ``smooth'' gauge to fix to.  Hence, we must alter the
definition of the usual constant-preserving \(P\) so that locally the
modes used in defining \(P\) form a basis for the low modes
of the system matrix, which for most simple pointwise smoothers are
also the modes not effectively treated.  This requirement, that a
small set of vectors partitioned into local basis functions can
approximate the entire lower end of the eigenspectrum of a matrix, is
known as the weak approximation property~\cite{Bramble:1991}.  It is
this property that leads to the success of
L\"{u}scher's~\cite{Luscher:2007se} deflation approach (where it is
referred to as local coherence) as well as our MG solver.

If the low modes are known, then the above MG process often yields
an optimal solver.  However, for the Wilson-Dirac system, these modes
are unknown and thus must be computed within the overall MG algorithm.
One viable approach, known in the MG literature as adaptive smooth
aggregation (\(\alpha\)SA)~\cite{Brezina:2004}, is given by
iteratively computing the low modes and then adjusting \(P\) to fit
them.  The general algorithm for computing these prototypes for a
given matrix \(A\) proceeds as follows.

In each adaptive step, the current solver\footnote{At the beginning of
  the setup, there exists no coarse grid, and so the current solver
  consists soley of the pre- and post-relaxation applications.} is
applied to the homogenous system, $Ax=0$, starting with a random
initial guess.  This tests the performance of the solver and also
produces a prototype of the slow-to-converge error.  At the $k$th step
of the adaptive process we obtain \(V^k = [ v_1, ..., v_k]\), with the
\(v_i\)'s denoting the computed prototypes.  As we iteratively augment
\(V^k\), we define the (tentative) prolongation operator \(P\) by
partitioning the candidate vectors into disjoint local blocks, and
compute a \(QR\) decomposition within each of these blocks.  The
global structure of the blocks, or {\it aggregates}, determines the
coarsening strategy.  The matrices \(Q\) form the columns of \(P\),
and \(R\) (of the $QR$ decomposition) represents the coefficients in
the coarse basis (\(V^k_c\)), i.e.,
\begin{equation}
P^{\dagger} P = I_c \hspace{3mm}\mbox{and}\hspace{3mm}P V^k_c = V^k\;.
\end{equation}
Whenever \(P\) is updated, the coarse operator is redefined to
complete the definition of the new solver.  The adaptive process
continues, iteratively augmenting \(V^k\), until convergence of the
evolving solver is deemed sufficient, say, for $k=N_v$ candidate
vectors.

\parasec{\bf Formulating an algorithm}

Generally, the two possible approaches for solving the non-Hermitian
Wilson system using MG are: (1) applying the adaptive MG approach to
the normal equations or (2) formulating the MG algorithm directly for
the Wilson-Dirac operator.

In the normal equations approach, the operator in question is HPD, and
hence variational MG convergence theory is applicable and the
two-level correction is optimal.  For the Dirac operator, however,
this approach increases the complexity of relaxation and the
coarsening.  In particular, the coarse operator \((D^\dagger D)_c =
P^\dagger (D^\dagger D) P\) does not involve only
nearest neighbor couplings, leading to loss in operator sparsity on
coarse levels.

The direct approach allows one to maintain a nearest neighbor coupling
among unknowns on the coarse level and, hence, to retain the sparsity
structure of the fine-level system.  Further, although the usual MG
convergence proofs generally do not apply, significant insight may be
obtained by considering the spectral decomposition of \(D =
|\psi_\lambda \rangle \lambda \langle \tilde\psi_\lambda| \), where
\(\psi\) and \(\tilde\psi\) are the right and left eigenvectors,
respectively, both having eigenvalue \(\lambda\). If we consider using
a Petrov-Galerkin oblique projection to deflate the eigenvector with
eigenvalue $\lambda$, then we have:
\begin{eqnarray}
  \mathcal{P} & = & \left(1 - D|\psi_\lambda\rangle\frac{1}{\lambda} \langle
  \tilde{\psi}_\lambda| \right) \\
& = & \left(1-D|\psi_\lambda\rangle\langle\tilde{\psi}_{\lambda'}|D|\psi_\lambda\rangle^{-1}
  \langle\tilde{\psi}_{\lambda'}|\right) \\
& \to & \left(1 - D P (R D P)^{-1} R
\right).
\end{eqnarray}
We thus see that prolongation should be defined using ``right null
space vectors'' and restriction using ``left null space vectors.''
Naively, this suggests that we define prolongation using smoothed
vectors of \(D\) and restriction from smoothed vectors of
\(D^\dagger\).  However, because of the \(\gamma_5\) symmetry of the
Wilson-Dirac operator, we have \(\tilde\psi_{\lambda^*} =
\gamma_5\psi_\lambda\) and, hence, a vector rich in low
right eigenvectors can be converted to one rich in low left
eigenvectors simply by multiplying by \(\gamma_5\).  

Given the current residual \(r_0\), our coarse-grid correction is thus given by
\begin{equation}
x_c =  P (P^\dagger \gamma_5 D P)^{-1} P^\dagger \gamma_5 r_0.
\end{equation}
Note that when coarsening the spin degrees of freedom together, the
coarse operator may have exactly zero eigenvalues.  As an example,
consider the free field operator, where the null space vector is constant.
Then, \(P^\dagger \gamma_5 P = 0\), and our
coarse-grid correction is ill-defined.  This can be avoided by keeping
chirality intact, i.e., by coarsening the upper and lower spin
components separately such that \(P^\dagger \gamma_5 = \sigma_3
P^\dagger\), where \(\sigma_3\) is the coarse space chirality matrix.
Hence, each prototype vector corresponds to two degrees of freedom on
the coarse lattice, and the \(\gamma_5\) factors cancel out in the
overall coarse-grid correction, yielding the former ``naive'' result
\(R = P^\dagger\).

The original adaptive smoothed aggregation approach introduced in
\cite{Brezina:2004} is essentially a black-box method, where the
coarsening strategy is chosen using an algebraic
strength-of-connection measure.  In lattice QCD, the system is
discretized on a uniform hypercubic lattice and the link matrices,
\(U_x^{\mu}\), belong to \(SU(3)\).  This motivates the use of
geometrically uniform coarsening.  The resulting coarse-grid operator
is nearest neighbor in spacetime, with effective link matrices of
dimension \(2N_v\times 2N_v\).  Recursing this coarsening procedure,
with the chiral components kept separate, maintains the sparsity
pattern and operator complexity on each of the successive levels.


With the prolongator and coarse-grid operator defined, all that
remains is to define a suitable relaxation procedure that effectively
damps the eigenvectors of the system matrix with eigenvalues that are
large in magnitude.  Classical MG methods use either Jacobi or
Gauss-Seidel smoothing, which are either inefficient in parallel or
cannot be applied directly to non-HPD operators.  We have found good
results using GMRES as a smoother (with under-relaxation parameter
\(\omega = 0.9\)); this yields a simple parallel approach that reduces
the residual in the \(D^\dagger D\) norm, ensuring that error
components corresponding to eigenvectors with large eigenvalues are
damped quickly.

Rather than being used as a stand-alone solver, MG is often employed
as a preconditioner to a Krylov process, thereby further accelerating
convergence.  The use of a non-stationary relaxation procedure (GMRES)
in our MG method requires that we use it as a preconditioner for an
appropriate {\em flexible} Krylov solver; here we used GCR(8) for the
Krylov solver.


\parasec{{\bf Numerical Results}} 
We have applied our MG-GCR solver directly to the Wilson-Dirac system
for a wide range of lattice spacings, gauge configurations and masses.
Our favored approach is to use \(4^4\) coarsening\footnote{The
  exception being where the lattice geometry restricts us to a less
  aggresive coarsening strategy, i.e., on the \(24^3\times64\) lattice
  we use \(4^4\) coarsening in moving from the fine grid to the first
  coarse grid, but \(2^3\times4\) from the first to second coarse
  grids.}  together with a 3-level V(0,4)-cycle, i.e., the
post-relaxation consists of the application of GMRES(4).  Furthermore,
a so-called W-cycle method is employed: for every correction to the
fine grid, two V-cycles are performed to update the intermediate grid.
On the coarsest grid, the system is solved using conjugate gradients
(CG) on the normal equations to a relative accuracy of \(10^{-3}\).
With these parameters we find that \(N_v = 20\) vectors is sufficient
to capture the null space of the Dirac operator, independent of the
lattice volume and lattice spacing.
\begin{center}
\begin{figure}[t]
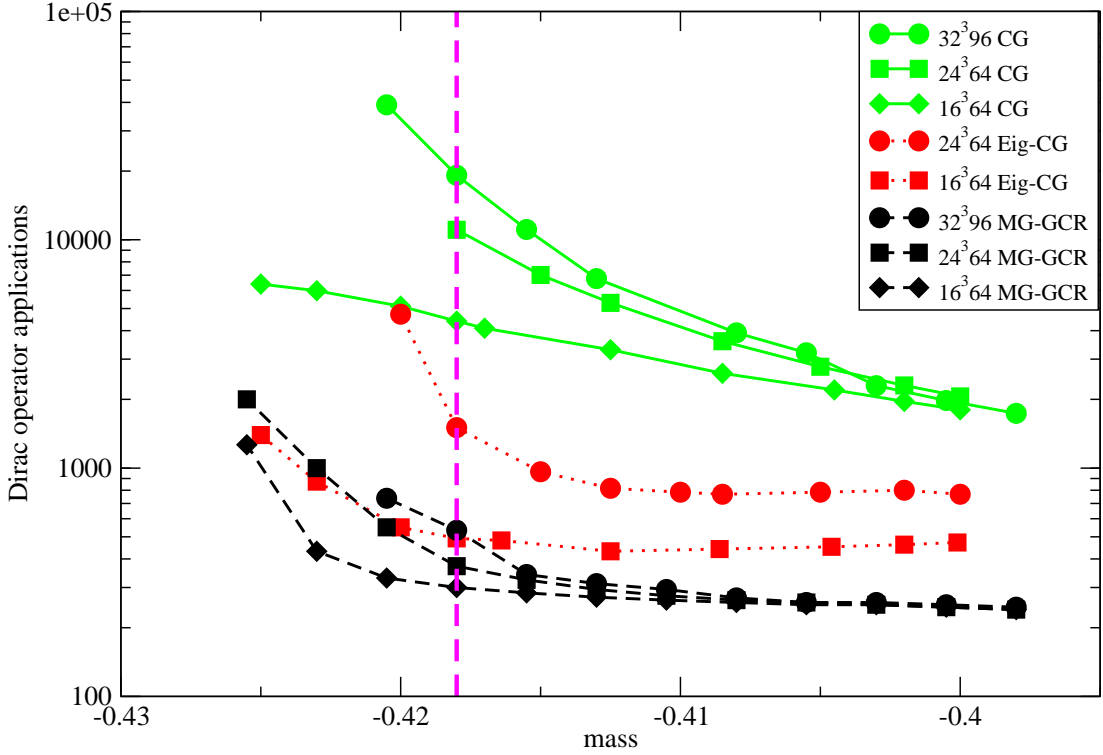

 \epspdffile{dirac_applications}
 \caption{Comparison of the total number of Wilson matrix-vector operations
   until convergence for CG, Eig-CG~\cite{Stathopoulos:2007zi} and
   MG-GCR (point sources, \(\beta = 5.5\), \(m_{crit} =
   -0.4175\), \(m_{sea}=-0.4125\), \(N_v = 20\) (MG-GCR), \(N_v=240\) (Eig-CG), outer
   solver tolerance \(=10^{-8}|b|\), gauge fields provided by the
  Hadron Spectrum Collaboration~\cite{Bulava:2009jb})}.
  \label{fig:dirac_applications}
\end{figure}
\end{center}

In Fig.~\ref{fig:dirac_applications}, we plot the total number of
Wilson-Dirac operator applications until convergence as a function of
fermion mass between red-black preconditioned CG, deflated CG (Eig-CG, results
adapted from \cite{Stathopoulos:2007zi}) and our MG-GCR algorithm for
three different volumes, where the lattice spacing and anisotropy have
been held fixed. For MG-GCR, this counts the work done on the fine
grid only.  It is evident that both Eig-CG and MG-GCR vastly reduce
the mass dependence that is seen with CG.  However, while MG-GCR
demonstrates close to ideal \(O(V)\) scaling over all three volumes,
the number of Eig-CG iterations approximately doubles from the
smallest to the intermediate volume.  Table~\ref{table:mg_iter} gives
the number of outer MG-GCR solver iterations for these same results,
clearly demonstrating the close-to-ideal scaling in both mass and
volume.  For both MG-GCR and Eig-CG, once the mass parameter
drops below the critical value that corresponds to zero physical fermion
mass (to the left of the vertical line), the prototypes /
eigenvectors no longer represent the null space of the operator, and
so the number of iterations increases rapidly.

\begin{table}[tb]
\begin{ruledtabular}
\begin{tabular}{cccc}
Mass & \(16^3\times64\) & \(24^3\times64\) & \(32^3\times96\) \\ \hline
-0.3980 & 40 & 40 & 41 \\
-0.4005 & 41 & 41 & 42 \\
-0.4030 & 42 & 42 & 43 \\
-0.4055 & 42 & 43 & 43 \\
-0.4080 & 43 & 44 & 45 \\
-0.4105 & 44 & 46 & 49 \\
-0.4130 & 45 & 49 & 52 \\
-0.4155 & 47 & 54 & 57 \\
\end{tabular}
\end{ruledtabular}
\caption{Number of iterations for the MG-GCR solver to reach convergence
  (parameters given in Fig.~\ref{fig:dirac_applications}). }
\label{table:mg_iter}
\end{table}

\begin{center}
\begin{figure}[tb]
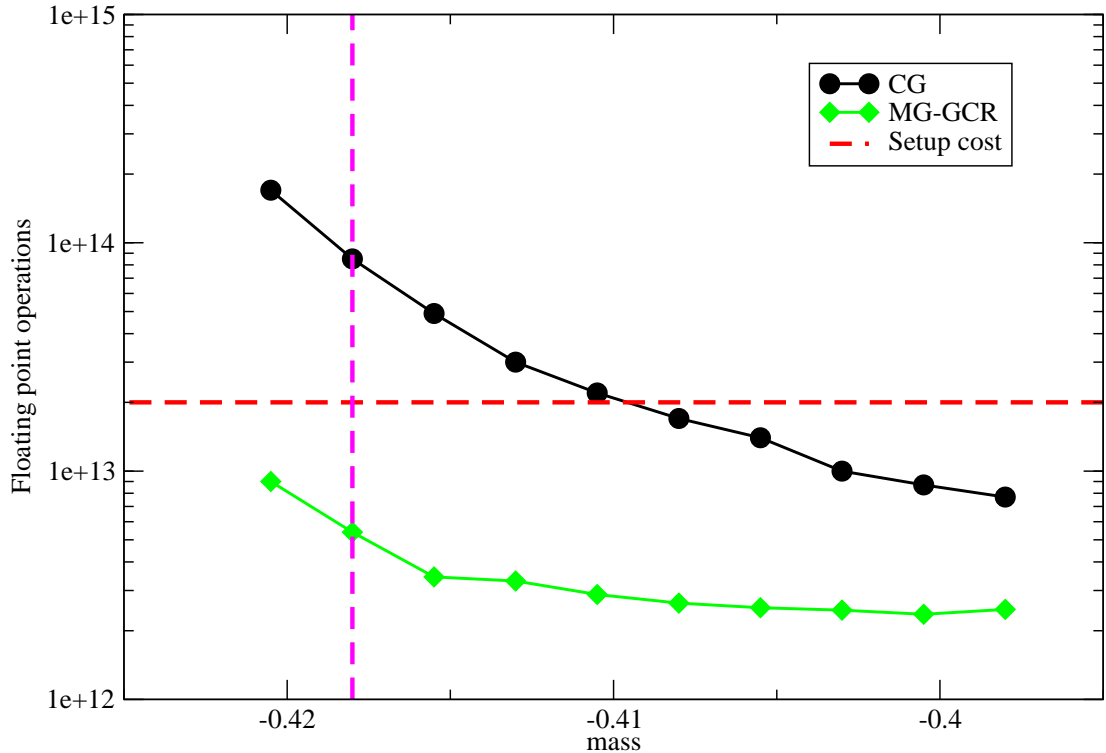

 \epspdffile{32_96_flops}
 \caption{Number of floating point operations required to reach
   convergence for CG and MG-GCR on the \(V=32^3\times96\) lattice
   (parameters given in Fig.~\ref{fig:dirac_applications}).  The
   horizontal line indicates the number of floating point operations
   required for the MG setup process.}
  \label{fig:flops}
\end{figure}
\end{center}

In terms of raw operation count, MG-GCR is comparable to Eig-CG on the
\(16^3\times64\) lattice, and 50\% more efficient on the
\(24^3\times64\) lattice.  In Fig.~\ref{fig:flops}, we plot the total
number of floating point operations to reach convergence on the
\(32^3\times96\) lattice for MG-GCR and CG.  It can be seen that the
use of multigrid reduces the total cost by a factor of three for heavy
quark masses, rising to a factor of 15 as the critical mass is
approached.

One important issue is the cost of the algorithm setup: the adaptive
process described above of sequentially finding prototypes to augment
\(V^k\) is expensive, since each prototype is found using the
then-current MG solver with \(k-1\) prototypes.  Noting that
relaxation alone will in practice yield a good initial guess for a
prototype, we instead adopt the following two-step process.  First,
we apply 10 iterations of relaxation to each of 20 random vectors to
define an initial \(V\).  We then divide the 20 resulting prototypes
into five groups of four and refine one group at a time by removing it
from \(V\) and iterating the truncated MG method five times upon the
prototypes in the group before reinserting it back into \(V\).  This
setup process need only be done at the critical mass (\(m =
m_{crit}\), \(\mbox{Re} (\lambda_{\tiny\mbox{{min}}}) \approx 0\)),
since the resulting null space representation can be used for all
heavier masses; this feature is independent of volume.  The setup cost
is equivalent to a single CG solve at an intermediate quark mass
(Fig.~\ref{fig:flops}).

\parasec{{\bf Concluding remarks}} 
In this work, we have introduced a new adaptive multigrid algorithm
for the non-Hermitian Wilson-Dirac operator.  The main results are the
near elimination of critical slowing down as the fermion mass is taken
to zero and the optimal scaling of the algorithm with volume.  These
developments promise to radically reduce the computational cost of
lattice field theory calculations.  Future work in this area will
focus on applying our algorithm in the context of full lattice QCD
simulations and developing these techniques for staggered and chiral
fermion discretizations of the Dirac operator.

{ \small This research was supported under: DOE grants
  DE-FG02-91ER40676, DE-FC02-06ER41440, DE-FG02-03ER25574 and
  DE-FC02-06ER25784; Lawrence Livermore National Laboratory contracts
  B568677, B574163 and B568399; and NSF grants PHY-0427646,
  OCI-0749202, OCI-0749317, OCI-0749300, DGE-0221680 and DMS-0810982. }

\end{document}